\documentstyle[12pt]{article}
\input epsf.sty
\newcommand{\bea}{\begin{eqnarray}}
\newcommand{\eea}{\end{eqnarray}}
\newcommand{\be}{\begin{equation}}
\newcommand{\ee}{\end{equation}}
\newcommand{\vs}[1]{\vspace{#1 mm}}

\renewcommand{\a}{\alpha}
\renewcommand{\b}{\beta}
\renewcommand{\c}{\gamma}
\renewcommand{\d}{\delta}

\newcommand{\dsl}{\pa \kern-0.5em /}

\newcommand{\pa}{\partial}

\newcommand{\nn}{\nonumber\\}

\begin{document}
\topmargin 0pt
\oddsidemargin 0mm

%\renewcommand{\thefootnote}{\fnsymbol{footnote}}

%\begin{titlepage}

\begin{flushright}

%USTC-ICTS-05-8\\

%MCTP-05-?\\

hep-th/0606041\\

%SINP-TNP/02-7

\end{flushright}
\vspace{6mm}

\begin{center}

{\Large \bf
Space-like branes, accelerating cosmologies and the near `horizon' limit}

\vs{10}

{\large Shibaji Roy\footnote{E-mail: shibaji.roy@saha.ac.in} and 
Harvendra Singh\footnote{E-mail: h.singh@saha.ac.in}}

 \vspace{6mm}

{\em
Saha Institute of Nuclear Physics,\\

 1/AF Bidhannagar, Calcutta-700 064, India}

\end{center}

\vspace{1cm}

\begin{abstract}
\begin{small}
It is known that there exist two different classes of time dependent solutions in 
the form of space-like (or S)-branes in the low energy M/string theory. Accelerating
cosmologies are known to arise from S-branes in one class, but not in the other 
where the time-like holography in the dS/CFT type correspondence may be more
transparent. We show how the accelerating cosmologies arise from S-branes in the
other class. Although we do not get the de Sitter structure in the lowest order
supergravity, the near `horizon' ($t\to 0$) limits of these S-branes are the 
generalized Kasner metric. 
\end{small} 
\end{abstract}
\newpage
\renewcommand{\a}{\alpha}
\renewcommand{\b}{\beta}
\renewcommand{\c}{\gamma}
\renewcommand{\d}{\delta}
\topmargin 0pt
\oddsidemargin 0mm
The interests in the time-dependent solutions in low energy (dimensionally reduced)
string/M theory are manifold. (a) They might lead to a better understanding of the
black hole as well as big bang singularities. (b) They might tell us about
how the cosmological observations of our universe can be understood from a 
fundamental theory. (c) The time dependent solutions can provide a concrete
realization of dS/CFT correspondence and show us how time emerges from an Euclidean
world-brane theory. (d) They can help us understand the time dependent processes
in string theory.

Low energy string/M theory admits various kinds of time dependent solutions. We
here consider a specific kind of solutions which are non-supersymmetric, singular
and has the form of space-like (or S)-branes 
\cite{Gutperle:2002ai}\footnote{See \cite{Lu:1996jk} for some 
earlier works on time dependent solutions in supergravities.}. 
An S$p$-brane has a $(p+1)$-dimensional
Euclidean world-volume and has isometry ISO($p+1$) $\times$ SO($d-p-2,1$)
in $d$ space-time dimensions. In the literature there exist two classes of S-brane
solutions. In one class the solutions are asymptotically ($t\to\infty$) non-flat and are 
characterized by three or more independent parameters \cite{Chen:2002yq,Ivashchuk:1997pk}. 
The coordinates used here are also
quite different from the usual BPS $p$-branes. Whereas, in the other class the
solutions are asymptotically ($t\to\infty$) flat and are characterized by two or 
more parameters \cite{Bhattacharya:2003sh}\footnote{In ten dimensions this class
of solutions was also obtained in \cite{Kruczenski:2002ap}.}. Here
the solutions can be written in terms of a single harmonic function with a
singularity at $t=0$ much like a BPS $p$-brane (of course the form
of the solutions are quite different) with the radial coordinate $r$ taking the 
place of time $t$. As the AdS/CFT correspondence is well-understood for BPS
D3-brane, it is hoped that for this class of solutions, the dS/CFT correspondence
\cite{Strominger:2001pn,Klemm:2001ea,Balasubramanian:2001nb}\footnote{See 
\cite{Park:1998qk} for some earlier works on dS/CFT correspondence.}
can similarly be understood by taking a near `horizon'\footnote{This is
really an abuse of the term ``horizon'', but we are using it in analogy with the
static, BPS $p$-brane solutions of M/string theory. Even for the latter case,
as is well-known, it is degenerate with zero area.} limit. However, since
it is difficult to find de Sitter solution in low energy string theory\footnote{
Some such solutions were found in \cite{Maeda:2004vm} with the inclusion of higher
order curvature terms in the effective action.}, the understanding of this 
issue remains unclear. The possibilities of obtaining  eternally accelerating
solutions from S-branes were also discussed in \cite{Forste:1998ed}.

One of our objectives here is to find the four dimensional cosmologies from the S-brane
solutions just mentioned. For this purpose let us recall that for the static
BPS $p$-brane solutions combining the radial part with the brane world-volume part 
we get a $(p+2)$ dimensional geometry which can be thought of as obtained by 
compactifying the ten dimensional theory on a $(8-p)$-dimensional sphere. In the
similar spirit, for S-branes, combining the temporal part with the $(p+1)$-dimensional
Euclidean brane world-volume part, we get a $(p+2)$-dimensional geometry by
compactifyting the $d$ dimensional theory on $(d-p-2)$ dimensional hyperbolic
space. So, in order to get a four dimensional world, we must take $p=2$. In general
the volume of compactification associated with the hyperbolic space is time dependent
and for this case it is possible to obtain accelerating cosmologies in four
dimensions, evading a `no-go' theorem \cite{gib,Maldacena:2000mw}, as was shown by 
Townsend and Wohlfarth \cite{Townsend:2003fx} starting 
from pure Einstein gravity in higher dimensions. The solution used in 
\cite{Townsend:2003fx} is 
a special case of the first class of S-brane solutions we mentioned above, that
is, they are asymptotically non-flat. In general in this case one should have a two
parameter family of solutions \cite{Chen:2002yq}, 
but ref.\cite{Townsend:2003fx} used some specific values of the 
parameters. A more general S-brane solutions containing a gauge field (and
a dilaton) and belonging 
again to the first class was used in refs.\cite{Ohta:2003pu,Roy:2003nd} 
to obtain four dimensional cosmology\footnote{Cosmological space-times from S-branes
were also studied earlier in ref.\cite{Burgess:2002vu}.}.
These solutions in general contains three (four) independent parameters and in showing
the accelerating cosmology the parameters are restricted to specific values. When
the gauge field (and the dilaton) is (are) put to zero these solutions reduce to 
those of Townsend and Wohlfarth when the parameters take specific values. 
Accelerating cosmologies were not known to follow
from the second class of S-brane solutions\footnote{In fact it might seem that
the accelerating cosmologies do not follow from this class of S-brane solutions.
The reason is, although it is known \cite{Roy:2002ik} that the first class of solutions 
maps to the
second class under a coordinate transformation on imposing the same boundary conditions 
on the metric and the dilaton but, for the parameter restrictions used in 
refs.\cite{Townsend:2003fx,Ohta:2003pu,Roy:2003nd} these two classes remains
disctinct. So, it may appear that the asymptotic non-flatness of the metric may 
be a necessary condition for the appearance of accelerating cosmologies. We will 
show later that this is not correct.}
where the coordinates are much like
those of BPS $p$-branes and dS/CFT type correspondence may be easily understood. 
So in this paper we will show that four dimensional accelerating cosmologies also follow
from the second class of S-brane solutions where the solutions are asymptotically
flat and are characterized by two or more independent parameters. 

The second class of $d$-dimensional S$p$-brane solutions were constructed in 
\cite{Bhattacharya:2003sh}.
They can also be obtained by a Wick rotation of the static, non-susy $p$-brane
solutions obtained in \cite{Lu:2004ms} as was shown there. Substituting $p=2$ 
and considering the 
$d-4=n$-dimensional hyperbolic space compactification 
\cite{Kaloper:2000jb,Starkman:2000dy,Kehagias:2000dg} on time varying volume
we obtain the four dimensional metric in the Einstein frame. We show that the
metric represents flat, homogeneous and isotropic FLRW universe with some scale 
factor $S(\eta)$. Then we show
that for certain values of the parameters, we can get an accelerating expansion
from the four dimensional metric which can be thought of as obtained from 
time dependent hyperbolic compactifications of low energy M or string theory for
$n=7$ or 6. It thus shows that the accelerating cosmologies quite generically follow
from the S-branes irrespective of the fact whether they are asymptotically flat
or not. Here also as for the other class of S-branes, the acceleration is transient
with two decelerating phases at $\eta \to 0$ and $\eta \to \infty$
and does not lead to a realistic cosmology. We find that as $\eta \to 0$, the
behavior of the scale factor is universal with $S(\eta) \sim \eta^{1/3}$ irrespective
of whether we consider first class or second class of S-brane solutions and whether
we have a gauge field and/or a dilaton or not. The reason for this can be attributed
to the fact that the near `horizon' or $\eta \to 0$ limit of these solutions give
the generalized Kasner metric\footnote{This was also noticed in a different context
in the second paper of 
reference \cite{Ivashchuk:1997pk} and \cite{Ivashchuk:1999rm}. We would like
to thank Vladimir D. Ivashchuk for pointing this out to us.}.      

The asymptotically flat, space-like or S2-brane solution in gravity coupled to
dilaton and an $(n-1)$-form gauge field in $n+4$ space-time dimensions is given
in \cite{Bhattacharya:2003sh} and has the form,
\bea
ds^2 &=& F^{\frac{12}{(n-1)\chi}}\left(H\tilde H\right)^{\frac{2}{n-1}}\left(
-dt^2 + t^2 dH_n^2\right) + F^{-\frac{4}{\chi}}\sum_{i=1}^3 (dx^i)^2\nn
e^{2\phi} &=& F^{-\frac{4 a(n+2)}{(n-1)\chi}}\left(\frac{H}{\tilde H}\right)^{2\delta}\nn
F_{[n]} &=& b {\rm Vol}(H_n)
\eea
In the above $F=\left(H(t)/\tilde {H}(t)\right)^\a \cos^2\theta + 
\left(\tilde{H}(t)/H(t)\right)^\b \sin^2\theta$, where, $H(t) = 1 + \omega^{n-1}/t^{n-1}$
and $\tilde{H}(t) = 1 - \omega^{n-1}/t^{n-1}$ are the two harmonic functions.
The physically acceptable region is $t>\omega$. Here
$\a$, $\b$, $\theta$, $\omega$ and $\d$ are integration constants. $b$ is a
`charge' parameter and $dH_n^2$ represents the line element of an $n$-dimensional
hyperbolic space and Vol($H_n$) is its volume form. $a$ is the dilaton coupling
which is given by $a^2 = 4 - 6(n-1)/(n+2)$ for maximal supergravities \cite{Duff:1993ye} 
in diverse
dimensions $d=n+4$. Note that $a=0$ for $n=7$ or for M-theory and $a=-1/2$ for
space-like D2-brane solutions of string theory.
Also in the above $\chi$ is
defined to be $\chi = 6 + a^2(n+2)/(n-1)$. Note that as $t \to \infty$, the functions
$H(t)$, $\tilde{H}(t)$ and $F$ go to unity and so the metric becomes asymptotically flat
in Rindler coordinates. We mentioned that the solution depend on several parameters
but not all of them are independent. They are related as
\bea
& & \a-\b \,\,=\,\, a\d\nn
& &\frac{1}{2}\d^2 + \frac{2\a(\a -a\d)(n+2)}{\chi(n-1)} \,\,=\,\, \frac{n}{n-1}\nn
& & b \,\,=\,\,\sqrt{\frac{4(n+2)(n-1)}{\chi}} (\a+\b) \omega^{n-1} \sin2\theta
\eea
So, the only independent paramaters are $\omega$, $\theta$ and $\delta$ ($\d$ vanishes
for $n=7$). 

Now instead of writing the solution in terms of two harmonic functions we
can rewrite it in terms of a single harmonic function much like the BPS $p$-branes.
For this purpose we make a coordinate transformation,
\be
\tilde {t} = t\left(1+ \frac{\omega^{n-1}}{t^{n-1}}\right)^{\frac{2}{n-1}} \,\,
=\,\, t H^{\frac{2}{n-1}}
\ee
This relation can be inverted to give
\be
t = \tilde{t} \left(\frac{1+\sqrt{f}}{2}\right)^{\frac{2}{n-1}}
\ee
where $f(\tilde{t}) = 1 - 4\omega^{n-1}/\tilde{t}^{n-1}$. The function $F(t)$
above can then be written as
\be
F = \left(\frac{H}{\tilde H}\right)^\a \cos^2\theta + \left(\frac
{\tilde H}{H}
\right)^\b \sin^2\theta = f^{-\frac{\a}{2}} \cos^2\theta + f^{\frac{\b}{2}} \sin^2\theta
\ee
The solution (1) can then be rewritten as,
\bea
ds^2 &=& F^{\frac{12}{(n-1)\chi}} f^{\frac{1}{n-1}}\left(
-\frac{d\tilde{t}^2}{f} + \tilde{t}^2 dH_n^2\right) 
+ F^{-\frac{4}{\chi}}\sum_{i=1}^3 (dx^i)^2\nn
e^{2\phi} &=& F^{-\frac{4 a(n+2)}{(n-1)\chi}} f^{-\delta}\nn
F_{[n]} &=& b {\rm Vol}(H_n)
\eea
By defining another coordinate
\be
\hat{t}^{n-1} = \tilde {t}^{n-1} - 4 \omega^{n-1}
\ee
we further rewrite the solution as
\bea
ds^2 &=& \left(g^{\frac{\a}{2}}\cos^2\theta + g^{-\frac{\b}{2}} \sin^2\theta\right)^
{\frac{12}{(n-1)\chi}} g^{\frac{1}{n-1}}\left(
-\frac{d\hat{t}^2}{g} + \hat{t}^2 dH_n^2\right)\nn 
& & \qquad\qquad\qquad\qquad\qquad
+ \left(g^{\frac{\a}{2}}\cos^2\theta + g^{-\frac{\b}{2}} \sin^2\theta\right)
^{-\frac{4}{\chi}}\sum_{i=1}^3 (dx^i)^2\nn
e^{2\phi} &=& \left(g^{\frac{\a}{2}}\cos^2\theta + g^{-\frac{\b}{2}} 
\sin^2\theta\right)^{-\frac{4 a(n+2)}{(n-1)\chi}} g^{\delta}\nn
F_{[n]} &=& b {\rm Vol}(H_n)
\eea
where $g(\hat{t}) = 1 + 4\omega^{n-1}/\hat{t}^{n-1}$. This solution has a singularity
at $\hat{t}=0$ much like the static, BPS $p$-brane which has singularity at $r=0$,
otherwise it is regular everywhere. The various parameters like $\a$, $\b$, $\omega$,
$\theta$, $\d$ and $b$ satisfy the same relations as given before in eq.(2). Also
in the following we will replace $\hat{t}$ in solution (8) by $t$ for brevity. As
mentioned before in order to obtain four dimensional cosmology we combine the temporal
part with the three dimensional Euclidean world-volume part and for obtaining
the four dimensional cosmology in Einstein frame we extract from the four dimensional
part the proper conformal factor. So, we write the metric in (8) as follows,
\be
ds^2 = F^{-\frac{6n}{(n-1)\chi}} g^{-\frac{n}{2(n-1)}}t^{-n} ds_E^2 + F^{\frac{12}
{(n-1)\chi}} g^{\frac{1}{n-1}} t^2 dH_n^2
\ee
where 
\be
ds_E^2 = -F^{\frac{6(n+2)}{(n-1)\chi}} g^{-\frac{n-4}{2(n-1)}}t^{n} dt^2
+F^{\frac{2(n+2)}{(n-1)\chi}} g^{\frac{n}{2(n-1)}}t^{n}\sum_{i=1}^3 (dx^i)^2
\ee
is the four dimensional metric in the Einstein frame. Here $F(t) = g(t)^{\a/2}
\cos^2\theta + g(t)^{-\b/2} \sin^2\theta$, with $g(t)$ as given before. Now
redefining a new time coordinate $\eta$ by
\be
d\eta = F^{\frac{3(n+2)}{(n-1)\chi}} g^{-\frac{n-4}{4(n-1)}} t^{\frac{n}{2}} dt
\ee
we can write the Einstein frame metric $ds_E^2$ in the standard FLRW form as
\be
ds_E^2 = -d\eta^2 + S^2(\eta)\left((dx^1)^2 + (dx^2)^2 + (dx^3)^2\right)
\ee
where the scale factor $S(\eta)$ has the form,
\be
S(\eta) \equiv A(t)= F^{\frac{n+2}{(n-1)\chi}} g^{\frac{n}{4(n-1)}} t^{\frac{n}{2}}
\ee
Let us also define another function
\be
B(t) = F^{-\frac{2(n+2)}{(n-1)\chi}} g^{\frac{n-2}{2(n-1)}}
\ee
We will have an expanding universe if the scale factor satisfies $dS(\eta)/d\eta > 0$
and the expansion will be accelerated if $d^2S(\eta)/d\eta^2 > 0$. In terms of the
functions $A(t)$ and $B(t)$, these two conditions can be written as
\bea
&& m(t) \equiv \frac{d \ln A}{dt} \,\, > \,\, 0\nn
&& n(t) \equiv \frac{d^2\ln A}{dt^2} + \frac{d\ln A}{dt} \frac{d\ln B}{dt} \,\, > \,\, 0
\eea
Now in order to understand whether (12) can give an accelerating cosmology or not
we will have to fix the various parameters and study the functions $m(t)$
and $n(t)$ of eqs.(15) to see
whether both can be satisfied for some range of the time coordinate
$t$ or $\eta$. We first note that from the second relation in eq.(2) we get $\a$
and $\b$ in terms of $\d$ as
\bea
\a &=& \pm \sqrt{\frac{\chi n - 3 \d^2(n-1)}{2(n+2)}} + \frac{a\delta}{2}\nn
\b &=& \pm \sqrt{\frac{\chi n - 3 \d^2(n-1)}{2(n+2)}} - \frac{a\delta}{2}
\eea
In the following we will study the four cases separately.
 
\vspace{.2cm}

\noindent (a) Pure Einstein gravity ($a=0$, $\d=0 \Rightarrow \a=\b$, $b=0 \Rightarrow
\theta=0$)\\
\noindent (b) Einstein gravity with an $(n-1)$-form gauge field ($\d=0 \Rightarrow \a=\b$,
$a=0$, $b \neq 0$)\\
\noindent (c) Einstein gravity with a dilaton ($a\neq 0$, $b=0 \Rightarrow
\theta=0$)\\
\noindent (d) Einstein gravity with a dilaton and an $(n-1)$-form gauge field.

\vspace{.2cm}

We will take $n=7$ for M-theory compactification (Case I) and $n=6$ for string theory
compactification (Case II). We will discuss the above four possibilities in these
two cases. Note that for $n=7$, there is no dilaton and so, $a=\d=0$
and therefore cases (c) and (d) do not arise. But for string theory all these four cases
will arise.

\vspace{.2cm}

\noindent {\it Case I: M-theory compactifications ($n=7$)}

\vspace{.2cm}

(a) For this case $a=0$, $\d=0$ and $\chi=6$ and so, $\a=\b=\pm\sqrt{7/3}$.
Also, we have $b=\theta=0$, so there is no gauge field. The
function $F=g(t)^{\a/2}$, so, from (13) and (14) we find 
\bea
A &=& g(t)^{\pm \frac{1}{8} \sqrt{\frac{7}{3}} + \frac{7}{24}}\, t^{\frac{7}{2}}\nn
B &=& g(t)^{\mp \frac{1}{4} \sqrt{\frac{7}{3}} + \frac{5}{12}} 
\eea
where $g(t) = 1 + 4\omega^6/t^6$.
We have plotted the functions $m(t)$ and $n(t)$ defined in (15) in Figures 1,2 
for the upper sign and have shown that both the conditions in (15) can be 
simultaneously satisfied for certain range of $t$ indicating an accelerating cosmology
for that period of time. The lower sign does not produce accelerated expansion. 
Note that we have plotted $m(t)$ and $n(t)$ for $4\omega^6 =
0.01,\, 0.1,$ and $1$ to show the behavior of the functions for different values
of the parameters. The parameter $4\omega^6$ has very little effect at very large and
very low values of $t$. This is the reason we see acceleration only for a finite
interval of time. When we introduce more and more parameters to study various
other M/string theory compactifications we will see that the basic behavior of the
functions remain the same and the additional parameters do not contribute 
significantly to the cosmology we obtain.

\begin{figure}[!ht]
\leavevmode
\begin{center}
\epsfysize=5cm
\epsfbox{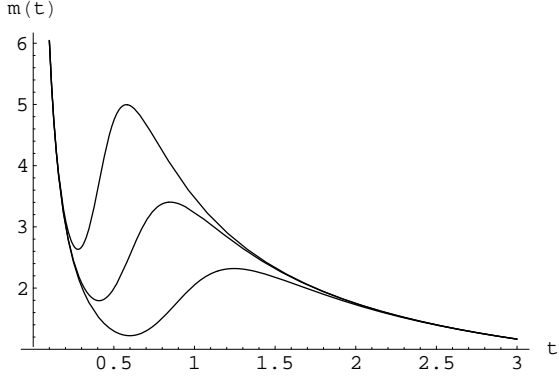}
\end{center}
\caption{\small \sl Plot of $m(t)$ for M-theory compactification without a gauge field
($\theta=0$) given in (15) for $4\omega^6 = 0.01,\,0.1,$ and $1$.
The top curve is for the value $0.01$, the middle one for $0.1$ and so on. All the
curves meet at small and large values of $t$.}
\end{figure}

\begin{figure}[!ht]
\leavevmode
\begin{center}
\epsfysize=5cm
\epsfbox{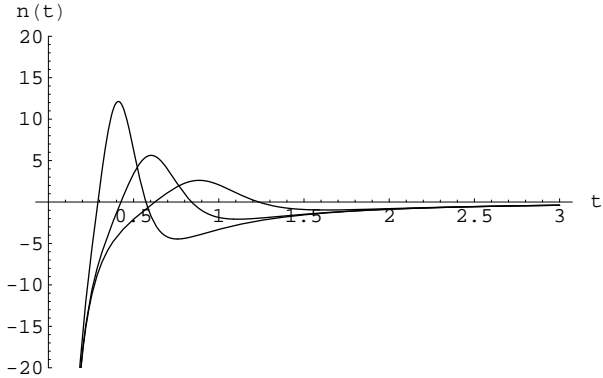}
\end{center}
\caption{\small \sl Plot of $n(t)$ for M-theory compactification without a gauge field
($\theta=0$) given in (15) for $4\omega^6 = 0.01,\,0.1,$ and $1$.
The left curve is for the value $0.01$, the middle one for $0.1$ and so on. All the
curves appear to meet at late and early times. All the curves are positive for some 
finite interval of time indicating an acceleration for that period, but are 
decelerating beyond that.}
\end{figure}

\vspace{.2cm}

(b) For this case $a=\d=0$ and so, $\a=\b = \pm\sqrt{7/3}$ as in the previous case.
However, here $b \neq 0$ which implies $\theta \neq 0$, that is, there is a non-zero
6-form gauge field. Here also $\chi=6$.
We have considered three different values of $\theta$ i.e. $\theta=\pi/4,\,\pi/6,$
and $\pi/18$. However we find that the functions behave almost similarly for
different values of $\theta$ and so, we give here the plots only for $\theta=\pi/4$.
In this case the various functions have the forms,
\bea 
F(t) &=& \frac{1}{2} \left(g(t)^{\frac{1}{2} \sqrt{\frac{7}{3}}} + 
g(t)^{-\frac{1}{2} \sqrt{\frac{7}{3}}}\right)\nn
A(t) &=& \frac{1}{2^{\frac{1}{4}}} \left(g(t)^{\frac{1}{2} \sqrt{\frac{7}{3}}} +
g(t)^{-\frac{1}{2} \sqrt{\frac{7}{3}}}\right)^{\frac{1}{4}} g(t)^{\frac{7}{24}}\, 
t^{\frac{7}{2}}\nn
B(t) &=& 2^{\frac{1}{2}} \left(g(t)^{\frac{1}{2} \sqrt{\frac{7}{3}}} +
g(t)^{-\frac{1}{2} \sqrt{\frac{7}{3}}}\right)^{-\frac{1}{2}} g(t)^{\frac{5}{12}}
\eea
where $g(t)$ is as given in (a) above. Here also we have plotted the functions
$m(t)$ and $n(t)$ given in (15) for $4\omega^6 = 0.01,\,0.1,\,1$ in Figures 3,4. 
Again the plots
show that we get an accelerating cosmology for certain interval of time.

\begin{figure}[!ht]
\leavevmode
\begin{center}
\epsfysize=5cm
\epsfbox{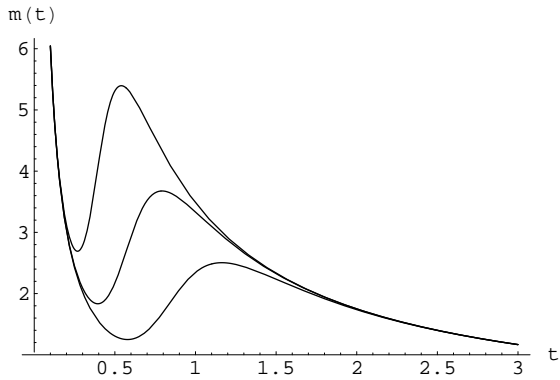}
\end{center}
\caption{\small \sl Plot of $m(t)$ given in (15) for M-theory compactification with
a gauge field for $4\omega^6 = 0.01,\,0.1,$ and $1$ and $\theta=\pi/4$.
The top curve is for the value $0.01$, the middle one for $0.1$ and so on. All the
curves meet at small and large values of $t$.}
\end{figure}

\begin{figure}[!ht]
\leavevmode
\begin{center}
\epsfysize=5cm
\epsfbox{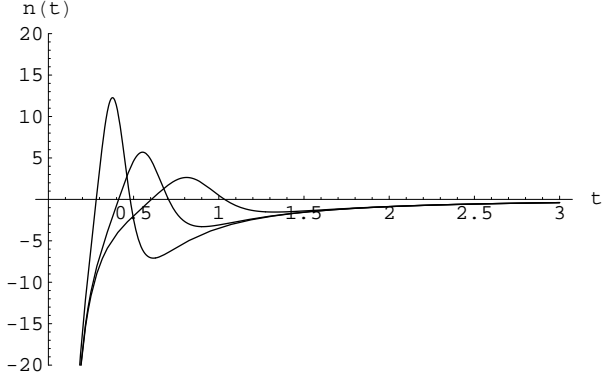}
\end{center}
\caption{\small \sl Plot of $n(t)$ for M-theory compactification with a gauge field
given in (15) for $4\omega^6 = 0.01,\,0.1,$ and $1$ and $\theta=\pi/4$.
The left curve is for the value $0.01$, the middle one for $0.1$ and so on. Here
$n(t)$ is positive in some finite time interval for different values of $4\omega^6$ 
chosen, showing the accelerating phase. But beyond these points there are deceleration.}
\end{figure}

\vspace{.2cm}

\noindent{\it Case II: String theory compactifications ($n=6$)}\footnote{Unlike
in cases (c), (d) cases (a), (b) given below do not correspond to maximal
supergravities.}

\vspace{.2cm}

(a) For this case $a=\d=0$, which implies $\a=\b=\pm 3/2$. Also $\chi=6$ and
$b=\theta=0$. The various functions in this case have the forms
\bea
F(t) &=& g(t)^{\pm\frac{3}{4}}\nn
A(t) &=& g(t)^{\pm\frac{1}{5}+\frac{3}{10}}\, t^3\nn
B(t) &=& g(t)^{\mp\frac{2}{5}+\frac{2}{5}}
\eea
where $g(t) = 1 + 4\omega^5/t^5$.
We have plotted $m(t)$ and $n(t)$ using the functions in (19). But since we have
pure Einstein gravity we expect the functions to have the same behavior as in M-theory
case or in Case I(a) above. We indeed find that to be true and so, we do not
give those plots here. Here also the upper sign of (19) gives accelerated expansion
but the lower sign gives deceleration.

\vspace{.2cm}

(b)  In this case we have $a=\d=0$ implying $\a=\b=\pm 3/2$. Here also
$\chi=6$, but now we have $b \neq 0$ which implies $\theta \neq 0$. We therefore
have a non-zero 5-form gauge field. The function $F$ has now the form
\be
F(t) = g(t)^{\pm \frac{3}{4}} \cos^2\theta + g(t)^{\mp \frac{3}{4}} \sin^2\theta
\ee
where $g(t)$ is as given in Case II(a) above.
We have plotted $m(t)$ and $n(t)$ for three different values of $\theta$
namely, $\theta=\pi/4$, $\theta=\pi/6$ and $\theta=\pi/18$ with three different
$4\omega^5 = 0.01,\,0.1,\,1$. Here also since this is pure gravity with a non-zero
gauge field, we expect the functions to behave similarly as in Case I(b) above.
We find this to be true and so, we do not give the plots here.

\vspace{.2cm}

(c) In this case we have a non-zero dilaton and so, $a=-1/2$. But we choose
$b=0$ which implies $\theta=0$ and the form field is vanishing. 
Here $\chi=32/5$ and from (16) we find
\bea
\a &=& \pm \sqrt{\frac{192-75\d^2}{80}} - \frac{\d}{4}\nn
\b &=& \pm \sqrt{\frac{192-75\d^2}{80}} + \frac{\d}{4}
\eea
The various functions defined in (13) and (14) now take the forms,
\bea
F(t) &=& g(t)^{\frac{1}{2} \left(\pm\sqrt{\frac{192-75\d^2}{80}} - \frac{\d}{4}
\right)}\nn
A(t) &=& g(t)^{\frac{1}{8} \left(\pm\sqrt{\frac{192-75\d^2}{80}} - \frac{\d}{4}\right) + 
\frac{3}{10}}\, t^3\nn
B(t) &=& g(t)^{\frac{1}{4} \left(\mp\sqrt{\frac{192-75\d^2}{80}} + \frac{\d}{4}\right) 
+ \frac{2}{5}}
\eea
where $g(t)$ is as given before in Case II(a) and $|\d| \leq \sqrt{192/75}$, 
but otherwise is an 
arbitrary parameter. We have plotted $m(t)$ and $n(t)$ as given in (15) for various
values of $\d$ and $4\omega^5$ in Figures 5,6. We found like the parameter 
$4\omega^5$ that for not all values of $\d$ we get
an accelerating cosmology. We chose a specific value of $\d=0.1$ and plotted the
functions for three different values of $4\omega^5$. For all these cases we get
accelerating cosmologies in some time interval. Here we have used only the upper 
sign of the various
functions given in (22). The lower sign gives deceleration. Since in this case
we have a non-zero dilaton, we expect
to have a different behavior of the functions. But surprisingly, the new parameter
$\d$ has little effect on the cosmology. The differences in behavior of the functions
are indeed very small as can be seen by comparing Figures 5,6 with those of the
M-theory compactifications given in Figures 1 -- 4.

\begin{figure}[!ht]
\leavevmode
\begin{center}
\epsfysize=5cm
\epsfbox{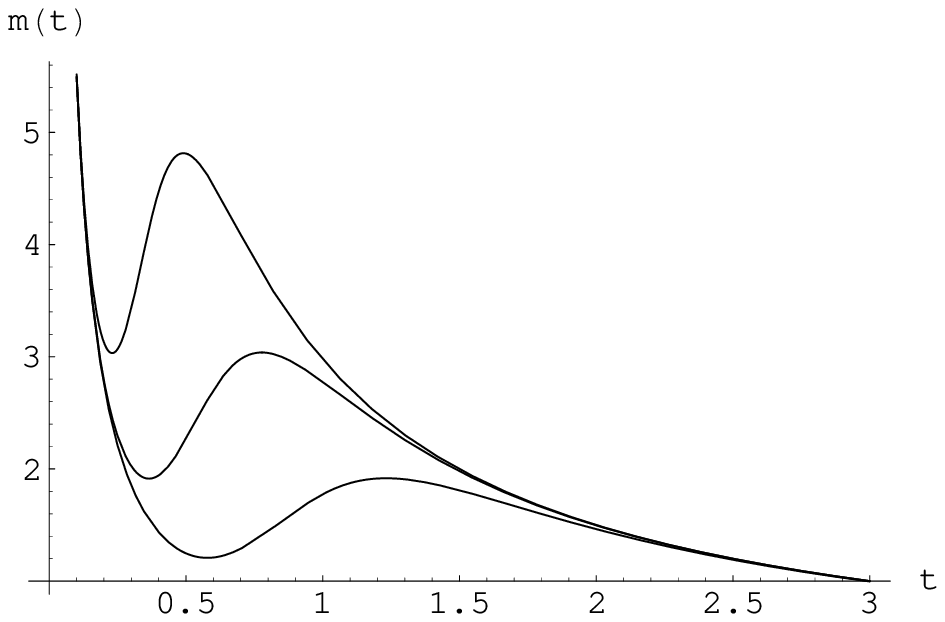}
\end{center}
\caption{\small \sl Plot of $m(t)$ given in (15) for  string theory compactification with
a dilaton but no gauge field ($\theta=0$) for $4\omega^5 = 0.01,\,0.1,$ and $1$ 
and $\d=0.1$.
The top curve is for the value $0.01$, the middle one for $0.1$ and so on. All the
curves meet at small and large values of $t$.}
\end{figure}
  
\begin{figure}[!ht]
\leavevmode
\begin{center}
\epsfysize=5cm
\epsfbox{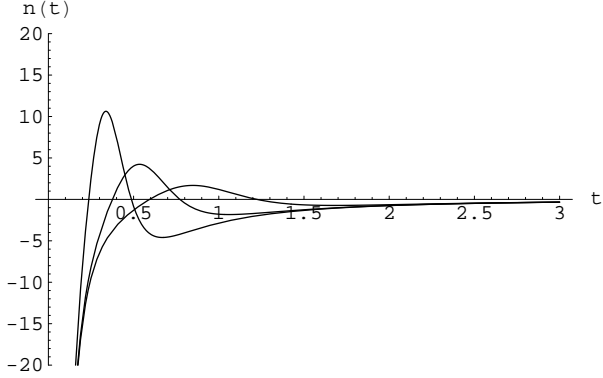}
\end{center}
\caption{\small \sl Plot of $n(t)$ given in (15) for  string theory compactification with
a dilaton but no gauge field ($\theta=0$) for $4\omega^5 = 0.01,\,0.1,$ and $1$ 
and $\d=0.1$.
The left curve is for the value $0.01$, the middle one for $0.1$ and so on. 
All the curves are positive for certain interval of time indicating an accelerating
cosmology. Beyond these points there are deceleration.}
\end{figure}
     
\vspace{.2cm}

(d) In this case none of the parameters are zero. So, we have a non-zero
dilaton as well as a non-zero 5-form gauge field. Again for this case $\chi=32/5$
and $a=-1/2$.
$\a$ and $\b$ are the same as in Case II(c) above. The various functions are
given as,
\bea
F(t) &=& g(t)^{\frac{1}{2} \left(\pm\sqrt{\frac{192-75\d^2}{80}} - \frac{\d}{4}
\right)} \cos^2\theta + g(t)^{\frac{1}{2} \left(\mp\sqrt{\frac{192-75\d^2}{80}} 
- \frac{\d}{4}\right)}\sin^2\theta\nn
A(t)&=& \left(g(t)^{\frac{1}{2} \left(\pm\sqrt{\frac{192-75\d^2}{80}} - \frac{\d}{4}
\right)} \cos^2\theta + g(t)^{\frac{1}{2} \left(\mp\sqrt{\frac{192-75\d^2}{80}}
- \frac{\d}{4}\right)}\sin^2\theta\right)^{\frac{1}{4}} 
g(t)^{\frac{3}{10}}\, t^3\nn
B(t) &=&  \left(g(t)^{\frac{1}{2} \left(\pm\sqrt{\frac{192-75\d^2}{80}} - \frac{\d}{4}
\right)} \cos^2\theta + g(t)^{\frac{1}{2} \left(\mp\sqrt{\frac{192-75\d^2}{80}}
- \frac{\d}{4}\right)}\sin^2\theta\right)^{-\frac{1}{2}}
g(t)^{\frac{2}{5}}
\eea
We have plotted the functions $m(t)$ and $n(t)$ given in (15) for various values
of $\theta$, $\d$ and $4\omega^5$. As before for not all values of the parameters 
we get accelerating cosmologies. We here give the plots for $\theta=\pi/6$, $\d=0.1$
and with three different values of $4\omega^5=0.01,\,0.1$ and $1$ in Figures 7,8. 
With more
parameters we expect the functions $m(t)$ and $n(t)$ to behave differently from the
other cases, but again we found that the additional parameters have very little effect
on the cosmology and we get essentially the same behavior as in the other cases. We
have plotted the functions in various cases with the same scale for better comparison.

\begin{figure}[!ht]
\leavevmode
\begin{center}
\epsfysize=5cm
\epsfbox{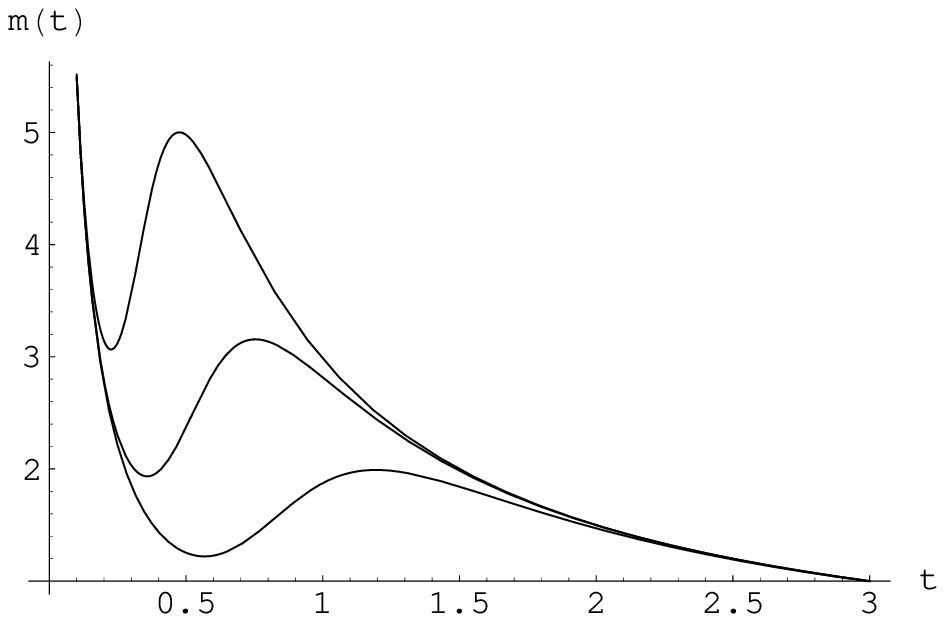}
\end{center}
\caption{\small \sl Plot of $m(t)$ given in (15) for  string theory compactification with
a dilaton and a gauge field for $4\omega^5 = 0.01,\,0.1,\,1$, $\d=0.1$ and $\theta=
\pi/6$.
The top curve is for the value $0.01$, the middle one for $0.1$ and so on. All the
curves meet at small and large values of $t$.}
\end{figure}

\begin{figure}[!ht]
\leavevmode
\begin{center}
\epsfysize=5cm
\epsfbox{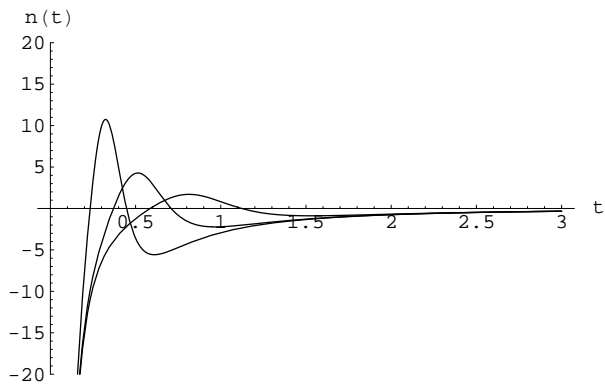}
\end{center}
\caption{\small \sl Plot of $n(t)$ given in (15) for  string theory compactification with
a dilaton and a gauge field for $4\omega^5 = 0.01,\,0.1,\,1$, $\d=0.1$ and $\theta
=\pi/6$.
The left curve is for the value $0.01$, the middle one for $0.1$ and so on. 
All the curves are positive for certain interval of time indicating an accelerating
cosmology. Beyond these points there are deceleration.}
\end{figure}

To summarize we have studied the conditions for accelerating cosmologies (15)
for various M/string theory compactifications on hyperbolic space with time varying
volume. In all the cases we have seen that both the conditions in (15) can be
simultaneously satisfied in a certain time interval for some specific values of the
various parameters. This indicates that under such compactifications the
resulting four dimensional isotropic, homogeneous, FLRW universe in the Einstein frame
can support accelerating cosmologies for certain period of time. This acceleration is
transient with an e-folding of the order of one and does not lead to an interesting
cosmological scenario. In order to obtain the accelerating cosmology we have made
use of the time dependent supergravity solution in the form of S-branes. A similar
phenomenon was known for a class of S-branes which are asymptotically non-flat 
\cite{Chen:2002yq,Ivashchuk:1997pk}.
However, here we have used another known class of S-branes which are asymptotically
flat \cite{Bhattacharya:2003sh,Kruczenski:2002ap}. Here we remark that although the 
details of the four dimensional cosmology we
obtain from these two classes are quite different, which are also seen from the
Figures given in this paper and those obtained in the earlier literature 
\cite{Townsend:2003fx,Ohta:2003pu,Roy:2003nd}, the 
robust features are quite similar. We here make a few comments regarding the various 
cases we have studied with the asymptotically flat solution. First of all, we have
given eight Figures, four for M-theory and four for string theory compactifications.
All the cases are different with different values of the parameters, yet the
Figures look quite the same. From all the plots of $m(t)$ we see that they remain
positive for all values of $t$ indicating that they give expanding universe. Also 
they have two extrema, one maxima
and one minima. The positions of these extrema and the variations of the curves
change significantly with the change of the parameter $4\omega^6$ (for M-theory)
or $4\omega^5$ (for string theory). But beyond certain time interval the function
$m(t)$ changes very little with $4\omega^6$ or $4\omega^5$. This is the reason the
plots $n(t)$ have two zeroes. So, the cosmologies start with deceleration (corresponding
to negative values of $n(t)$) and also end in deceleration, but since they
have two zeroes we have accelerating phase in between. However,
since the positions of extrema changes with $4\omega^6$ or $4\omega^5$, the
period of acceleration also changes with the parameter. But we must remark
that the peroid of acceleration does not change much with the introduction of
other parameters (like $\d$ and/or $\theta$). It is easy to see that the relation
between the actual four-dimensional time $\eta$ and $t$ given in eq.(11) can
be integrated for large ($t \to \infty$) and small ($t \to 0$) $t$. For $t \to \infty$
we find $t \sim (\eta -\eta_0)^{2/(n+2)}$ and so, from (13) we get the late time
behavior of the scale factor as $S(\eta) \sim (\eta-\eta_0)^{n/(n+2)}$. So, the
behavior depends on the theory from which the cosmology is obtained, but the
detailed structure is not important. However, keeping the upper signs of $\a$
and $\b$ given in (16), it is easy to check that
at early time $F(t) \sim t^{-(n-1)\a/2}$ and $t \sim \eta^{-4\chi/(6\a(n+2) 
- 3n\chi)}$. Similarly for the lower signs of $\a$ and $\b$, we can check
that at early time, $F(t) \sim t^{(n-1)\b/2}$ and $t \sim \eta^{4\chi/(6\b(n+2)
+ 3n\chi)}$. For both cases 
the behavior of the scale factor at early time can be seen from (13)
to have the form $S(\eta) \sim \eta^{1/3}$. Now we find that the behavior is
quite universal independent of $n$. These early and late time behaviors of the
scale factor match exactly with the pure gravity compactification (although
their solution is different from the one used here) observed by
Townsend and Wohlfarth \cite{Townsend:2003fx}. 

The universal feature of the scale factor at early time
can be attributed to the fact that the metric takes the generalized Kasner form
when $t \to 0$ and the resulting space-time is four dimensional. We would like to
remark that the solution we have used in this paper is asymptotically flat
with a `horizon' at $t=0$, just like the static, BPS $p$-brane (with horizon at
$r=0$) and since the near horizon geometry of a static D3-brane has the anti 
de Sitter structure, similarly, one might expect to get a de Sitter structure
for this time dependent S2-brane solution in the near `horizon' limit. But, it 
can be easily
checked for the solution (8) that, we get de Sitter structure
only if $\d$ is a complex number or in other words a real de Sitter solution does
not follow from
the near `horizon' limit of S2-brane. However, we find that $t\to 0$ 
limit of the metric
and the dilaton in (8) have the forms\footnote{Here we keep only the upper signs 
of $\a$ and $\b$ given in (16) and mention about the lower signs later. Also
since for very low values of $t$, $4\omega^{n-1}$ has very little effect on cosmology
we have put $4\omega^{n-1}=1$ without any loss of generality.},
\bea
ds^2 &=& -t^{-\frac{6\a}{\chi} + n -2} dt^2 + t^{\frac{2\a}{\chi}(n-1)} \sum_{i=1}^3
(dx^i)^2 + t^{-\frac{6\a}{\chi}+1} dH_n^2\nn
e^{2\phi} &=& t^{\frac{2 a\a(n+2)}{\chi} - \d (n-1)}
\eea
Since in the above $t$ is nearly zero, one can rewrite the metric by 
replacing $t$ by $\lambda\,t$
with $\lambda \to 0$ and $t$ = finite, in the form
\be
ds^2 = \lambda^{-\frac{6\a}{\chi}+n}\left[-t^{-\frac{6\a}{\chi} + n -2} dt^2 + 
\lambda^{\frac{2\a}{\chi}(n+2) -n} t^{\frac{2\a}{\chi}(n-1)} \sum_{i=1}^3
(dx^i)^2 + \lambda^{1-n} t^{-\frac{6\a}{\chi}+1} dR_n^2\right]
\ee 
Note that we have replaced $dH_n^2$ in (24) by the line element of a flat
$n$-dimensional Euclidean space $dR_n^2$. This is possible because for $n>1$,
the coefficient of $dH_n^2$ in (24) becomes infinite and so we can replace 
it by the flat space. Going back to the original coordinate $t$ we therefore write 
the solution (24) by replacing $dH_n^2$ by $dR_n^2$. 
Now defining a new coordinate by $d\bar{t} = t^{-(3\a/\chi) + (n/2) -1} dt$ we can 
write the metric and the dilaton in (24) as,
\bea
ds^2 &=& - d\bar{t}^2 + \bar {t}^{-\frac{4\a(n-1)}{6\a - \chi n}} \sum_{i=1}^3
(dx^i)^2 + \bar {t}^{\frac{2(6\a-\chi)}{6\a-\chi n}} dR_n^2\nn
e^{2\phi} &=& \bar{t}^{\frac{-4 a\a(n+2) +2\d \chi (n-1)}{6\a - \chi n}}
\eea    
Further, defining the various exponents of $\bar {t}$ appearing in the metric and
the dilaton in (26)
\be
p = -\frac{2\a(n-1)}{6\a-\chi n}, \qquad q = \frac{6\a-\chi}{6\a - \chi n}, \qquad
\gamma = \frac{-2 a\a(n+2) + \d \chi (n-1)}{6\a - \chi n}
\ee
we find that they satisfy
\be
3p + n q = 1, \qquad 1 - 3p^2 - n q^2 = \frac{1}{2} \gamma^2
\ee 
In order to satisfy the second equation of (28) we have made use of the relation
between the parameters $\a$ and $\d$ given in (2). These are precisely the conditions
satisfied by the generalized Kasner metric \cite{Das:2006dz}. Note that in deriving
(24) and (26) we have used the upper signs of $\a$, $\b$, however, for the lower signs,
they have very similar forms with $\a$ replaced by $-\b$. The various exponents
of $\bar {t}$ in that case have the forms
\be
p = -\frac{2\b(n-1)}{6\b+\chi n}, \qquad q = \frac{6\b+\chi}{6\b + \chi n}, \qquad
\gamma = -\frac{2 a\b(n+2) + \d \chi (n-1)}{6\b + \chi n}
\ee 
It is easy to check that they again satisfy eq.(28) if we use eq.(2).
We thus conclude that the near `horizon'
limit of the solution (8) is the generalized Kasner metric. Note that when there
is no dilaton $\gamma=0$, $a=0$ and $\d=0$, then $p$ and $q$ reduce precisely to
the standard Kasner form given in \cite{Townsend:2003fx}. Because of this 
the scale factor
of the resulting four dimensional FLRW cosmology has a universal behavior
$S(\eta) \sim \eta^{1/3}$.

We have thus seen that the four dimensional accelerating cosmologies
quite generically follow from the known S-brane solutions of M/string theory.
This was known for the asymptotically non-flat solutions and we have shown
this to be true for the asymptotically flat solutions as well. As in the former
case, we have seen that the cosmology does not change much with
the introduction of various parameters. Although we get accelerating cosmologies
in various cases, the acceleration is transient. Since S-branes are unstable
systems without any supersymmetry, it might be interesting to see whether by coupling
the four dimensional action with the tachyon effective action and/or the
inclusion of the higher order curvature terms can give
a better understanding of the initial cosmological singularity and a longer
period of acceleration leading to a de Sitter solution. 

\vspace{.5cm}

\noindent {\bf Acknowledgements}

\vspace{2pt}

H.S. is grateful to AS-ICTP, Trieste and its Associateship Programme for
hospitality where part of this work has been carried out.

\end{document}